\documentclass{revtex4}
\usepackage{amssymb}
\usepackage{amsmath}
\usepackage{xcolor}
\usepackage{url}
\usepackage{graphicx}  % Para incluir imagens
\usepackage[utf8]{inputenc}
\usepackage{xcolor}
\usepackage[utf8]{inputenc}
\usepackage{amsmath,amssymb} 
\usepackage{epsfig}   
\usepackage[english]{babel}

\begin{document}

\renewcommand{\thefootnote}{\fnsymbol{footnote}} 

\title{Scaling of Street Network Centrality with City Population} 
\author{R. L. Fagundes$^{a,1}$, G. G. Piva$^a$, A. S. Mata$^{a}$, F. L. Ribeiro$^{a,} $ \footnotemark{} }
\affiliation{$^a$ Universidade Federal de Lavras, Departamento de Física - UFLA, Lavras, MG, Brazil}
\affiliation{$^1$ IOER, Weberplatz 1 01217 Dresden, Germany.}
\footnotetext[1]{Corresponding author: \url{fribeiro@ufla.br}}

\begin{abstract}

Urban scaling laws reveal how cities evolve as their populations grow, yet the role of street network accessibility in this process remains underexplored. We analyze over 5,000 Brazilian cities to establish a scaling law linking average closeness centrality $\langle c_C\rangle$—a measure of structural accessibility in street networks—to population size $N$. Our results demonstrate that $\langle c_C \rangle$ decays sublinearly as $N^{-\sigma}$ ($\sigma \approx 0.38$), indicating that larger cities redistribute accessibility from cores to peripheries while maintaining navigability through hierarchical shortcuts. This scaling arises from the fractal interplay between infrastructure and population, characterized by a network dimension $d \approx 2.17$, which exceeds that of a 2D grid. The slower decline in closeness centrality ($\sigma <0.5$) reflects a trade-off: urban expansion reduces proximity but enhances connectivity through optimized path diversity, fostering economic dynamism. By integrating the Molinero \& Thurner model with network centrality metrics, we provide a framework to reconcile infrastructure efficiency with equitable accessibility in growing cities.


\end{abstract}

\maketitle

\section{Introduction}
\label{sec1}
Cities are hubs of social, economic, political, and cultural activity, where interactions between individuals and their environment shape emergent urban dynamics. These interactions drive socioeconomic output, with road networks forming the physical backbone that enables their propagation across urban scales. Crucially, while both human interactions and infrastructure underpin urban efficiency, they exhibit divergent scaling behaviors with respect to city population size \cite{Bettencourt2007, Ribeiro2017, Meirelles2018, Marc2019, Dong2020, Molinero2021, Rybski2022}.

Empirical studies reveal that many urban metrics, from social interactions \cite{schlapfer2015} to road network properties \cite{Molinero2021}, scale with population size $N$ via a power law:
\begin{equation}
Y \sim N^{\beta},
\label{eq:scaling_law }
\end{equation}
where the exponent $\beta$ distinguishes between regimes of increasing returns ($\beta>1$, e.g., GDP, innovation), economies of scale ($\beta<1$, e.g., road networks), and linearity ($\beta=1$, e.g., water consumption) \cite{Bettencourt2007, strano2016rich, Molinero2021}. These scaling patterns reflect fundamental principles of complex systems: cities self-organize through non-trivial agent interactions, exhibit multifractal properties in infrastructure and population distributions \cite{Murcio2015, Zhou2020, Stepinski2020}, and resemble dissipative systems exchanging energy and matter with their environment \cite{Hidalgo2021, Broekel2023}. Theoretical frameworks link these dynamics to mechanisms such as mobility-productivity trade-offs, fractal infrastructure, and rank-size distributions \cite{Bettencourt2013, Yakubo2014, Molinero2021, Ribeiro2023}, offering insights into mobility patterns, policy design, and infrastructure resilience.

Within this context, street networks play a dual role: they enable socioeconomic interactions while embodying hierarchical complexity. Vehicular flows concentrate unevenly across streets, reflecting structural hierarchies that influence accessibility, multimodal transport use, and emissions \cite{Lammer2006, Bassolas2019}. Centrality measures, particularly closeness centrality, quantify this hierarchy by identifying streets critical to connectivity and urban function \cite{Freeman1977}. Closeness centrality, for instance, correlates with public-oriented activities like retail services \cite{porta2009correlating} and informs resilience analyses, as disruptions to central roads disproportionately impact traffic \cite{Robson2021, Nair2019}. Yet, despite its relevance, a scaling law linking average closeness centrality to population size remains unestablished—a gap limiting our understanding of how infrastructure adapts to population growth and complexity.

This study addresses this gap by analyzing street networks in over 5,000 Brazilian cities to derive an empirical scaling law for average closeness centrality. Building on Molinero \& Thurner’s model \cite{Molinero2021}, we propose a theoretical framework connecting population-driven infrastructure demands to structural accessibility. Our results deepen the interplay between scaling laws and network science, offering insights into urban adaptation and informing equitable infrastructure planning.

\section{Materials and Methods}
\label{sec2}

The study leverages data from the IBGE API, including 2022 population estimates from Brazil’s most recent census \cite{population2022}. Street networks for over 5,000 Brazilian cities were extracted using the \texttt{osmnx} Python package \cite{OpenStreetMap}, with urban boundaries defined by satellite-derived 2015 urban area data \cite{embrapa2015}. This combination ensures spatial alignment between population metrics and infrastructure topology.

Each city’s road network is modeled as an undirected, weighted spatial graph $G=\{ E,V\}$, where $E$ represents street segments (edges) and $V$ denotes intersections or endpoints (nodes). Nodes are embedded in 2D space via coordinates $(x,y)$, preserving geographic fidelity. For each city, $n=|V|$ and $e=|E|$ define the graph’s size and connectivity.

Shortest path distances $\ell_{ij}$ between nodes $i$ and $j$ underpin key metrics. The farness of node $i$ is $f_i=\sum_{i\neq j} \ell_{ij}$, and its closeness centrality is $c_i=1/f_i$, quantifying proximity to all other nodes. The network’s average shortest path length $\ell$ scales with $n$ in a $d$-dimensional lattice as $\langle \ell \rangle \propto n^{1/d}$ \cite{mata2020}, implying dimensionality-dependent accessibility.

Formally, closeness centrality is defined as \cite{murray1965}:
\begin{equation}
c_{C}(i) = (n-1) \sum_{i\neq j}\frac{1}{\ell_{ij}},
\label{fig:closeness_centrality_expression }
\end{equation}
where $n-1$ normalizes by reachable nodes.   Particularly in planned cities, the average closeness $\langle c_C \rangle \propto 1/ \langle \ell \rangle$ thus scales as:
\begin{equation}
\langle c_C \rangle \sim n^{-1/d}.
\end{equation}
This scaling reflects the network’s spatial efficiency, with lower $d$ (e.g., $d=2$ for planar grids) correlating with slower decay in accessibility.

Closeness centrality has dual interpretive value: it signals transportation cost savings at locations (nodes) \cite{Sevtsuk2012} and predicts facility distributions (e.g., residential, retail) \cite{shi2024}. Originating in social network analysis \cite{bavelas1950}, its adaptation by Murray and Freeman emphasized efficiency and interdependence in spatial systems \cite{murray1965, freeman1978}.

\begin{figure}[!htbp]
    \centering
\includegraphics[scale=0.25]{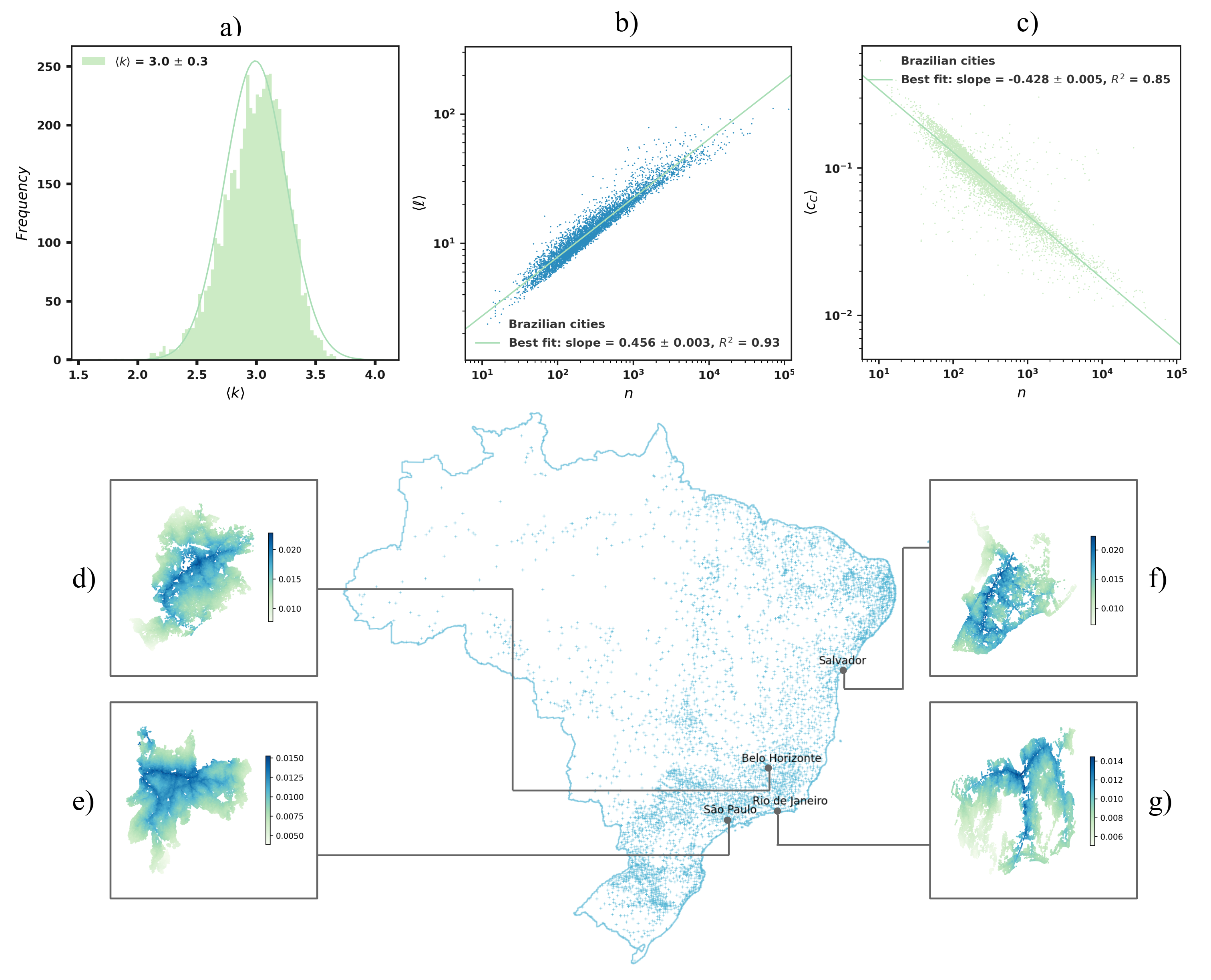}
    \caption{The average degree a),  scaling relations between nodes and average shortest path with a positive slope of $0.456\pm0.003$ b), and average closeness centrality with a negative slope of $-0.428\pm 0.005$ c). Street networks with closeness centrality heatmaps for four Brazilian cities: d) Belo Horizonte (MG), e) São Paulo (SP), f) Salvador (BA), and g) Rio de Janeiro (RJ).}
    \label{fig:infographic}
\end{figure}

\section{Results}
\label{sec3}

The histogram in Figure~\ref{fig:infographic} a) reveals that Brazilian cities exhibit quasi-regular street networks, with an average node degree $\langle k \rangle = 3\pm 0.3$. This suggests a semi-planned urban structure where most intersections connect three streets. Figure~\ref{fig:infographic} b) demonstrates a power-law scaling between the average shortest path length $\langle \ell \rangle$ and the number of nodes $n$, expressed as $\langle \ell \rangle \sim n^{1/d}$. The derived network dimension $d \approx 2.17$ (vs. $d = 2$ for a regular grid) indicates enhanced navigability, as path lengths grow slower than in strictly planar lattices.  

This deviation from $d = 2$ arises from occasional high-degree nodes that act as shortcuts, reducing overall connectivity distances. Compared to UK street networks ($\langle k \rangle \approx 2.4$ \cite{Arcaute2016}), Brazilian networks balance grid-like regularity with limited randomness, avoiding the logarithmic scaling ($\langle \ell \rangle \sim \ln n$) typical of random graphs.
\begin{figure}[!h]
    \centering
    \includegraphics[scale=.8]{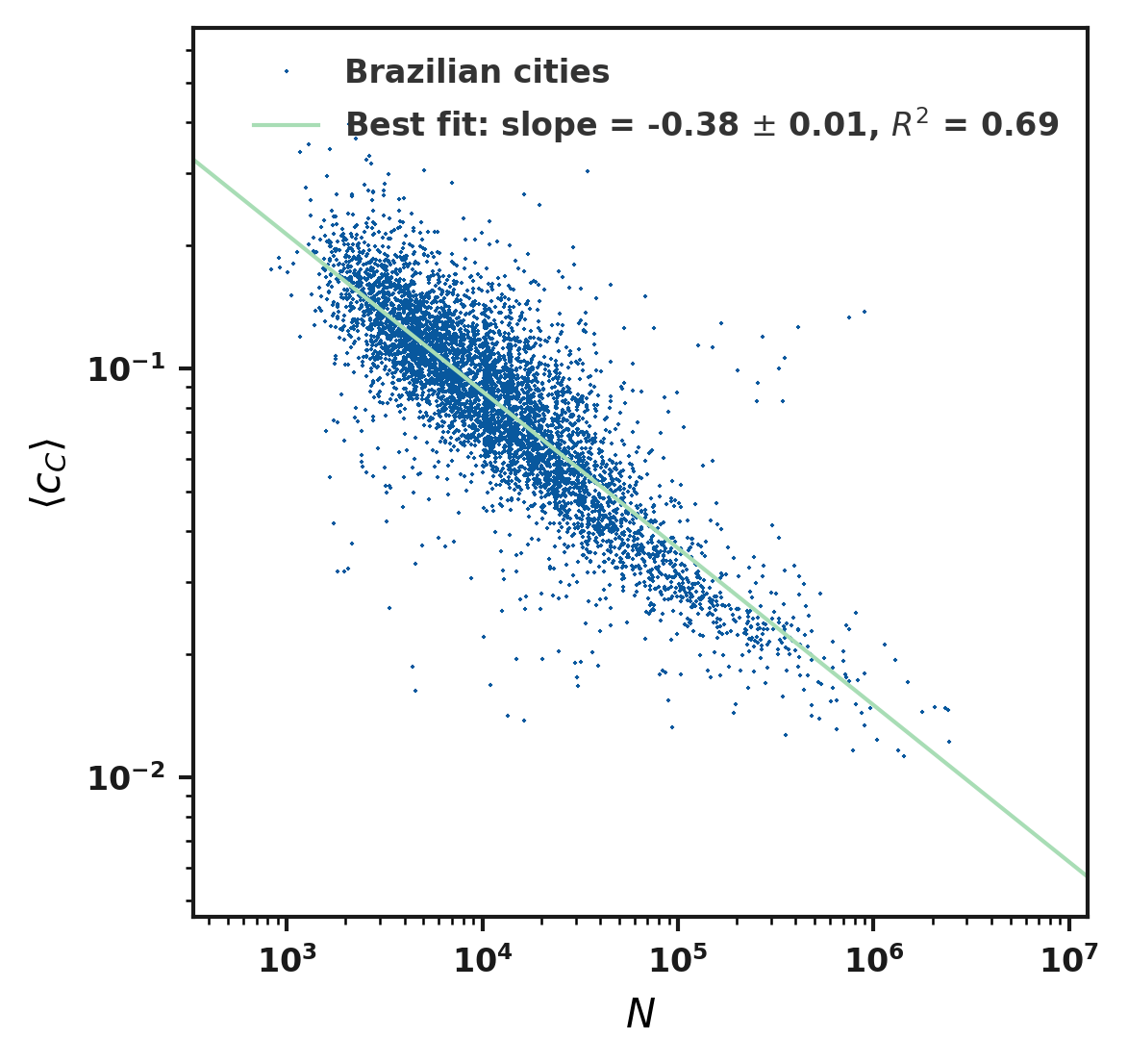}
    \caption{Empirical scaling of average closeness centrality $\langle c_C \rangle$ versus population $N$. The green line denotes the power-law fit $\langle c_C \rangle \sim N^{-\sigma}$.}
    \label{fig:emp_clos_centr}
\end{figure}

On the other hand, closeness centrality $\langle c_C \rangle$ declines with $n$ as $\langle c_C \rangle \sim n^{-1/d}$ (Fig.~\ref{fig:infographic} c), signaling a gradual redistribution of accessibility from central to peripheral areas. Visualizations of four cities (Figs.~\ref{fig:infographic} d-g) confirm that nodes in central urban zones exhibit higher closeness centrality, aligning with heatmap patterns of proximity.  

Figure~\ref{fig:emp_clos_centr} establishes the empirical scaling of $\langle c_C\rangle$ with population $N$:  

\begin{equation}
    \langle c_C \rangle \sim N^{-\sigma}, \quad \sigma = 0.38 \pm 0.01.
\end{equation}

This power-law decay reflects reduced proximity between urban centers and peripheries as cities expand. Theoretically, this aligns with Molinero \& Thurner’s model \cite{Molinero2021}, where street length $L \propto N^{\beta_{\text{sub}}}$ ($\beta_{\text{sub}} = d_f/d_p < 1$, where $d_f$ and $d_p$ are infrastructure and population fractal dimensions, respectively.) and $L \propto n$. Substituting into $\langle c_C \rangle \sim n^{-1/d}$ yields:  
\begin{equation}
\langle c_C \rangle \propto N^{-\beta_{\text{sub}}/d},
\label{eq:c_C}
\end{equation}  
consistent with our empirical $\sigma$. The reduced rate of decay ($\sigma = 0.38$ for $d = 2.17$) emphasizes the mitigating effect of shortcuts in preserving accessibility during growth, which results in an exponent $\beta_{\textup{sub}}=0.83$, as observed in the reference~{\cite{Molinero2021}.

\section{Discussion and Conclusion}

This study establishes an empirical scaling law linking average closeness centrality $\langle c_C \rangle$ in street networks to city population $N$, building on Molinero \& Thurner’s framework for infrastructure scaling \cite{Molinero2021}. The observed power-law decay $\langle c_C \rangle \sim N^{-0.38}$ signals that urban growth reduces average proximity between nodes, necessitating longer travel paths as networks expand. Critically, the sublinear exponent ($\sigma \approx 0.38$)—smaller than the theoretical $0.5$ for a 2D grid—reflects compensatory mechanisms in network organization. These include the interplay between the fractal dimensions ($d_f$) and ($d_p$), as well as hierarchical street layouts that mitigate accessibility loss through shortcuts \cite{fang2021spatial, song2023interactive, maniadakis2013structural}.  

The derived relationship $\langle c_C \rangle \propto N^{-\beta_{\text{sub}}/d}$ (Eq.~\ref{eq:c_C}) underscores how sublinear infrastructure scaling ($\beta_{\text{sub}} < 1$) and network dimensionality ($d \approx 2.17$) jointly shape accessibility. This highlights a trade-off: while larger cities require proportionally less road infrastructure per capita, their spatial complexity redistributes centrality from cores to peripheries, fragmenting accessibility. Urban planners must balance this duality—leveraging economies of scale in infrastructure while preserving equitable access through hierarchical design.  

Future work should integrate fractal geometry with network science to refine scaling models, particularly how population density gradients and verticalization alter transportation accessibility. Additionally, our findings open avenues to study commercial land-use dynamics, as retail activity correlates with closeness centrality \cite{porta2009correlating}. Empirical limitations, such as sparse data on global street networks and population fractality, call for expanded datasets to validate universality across urban typologies. 

\section*{Author contributions: CRediT}
\textbf{R. L. Fagundes}: Conceptualization, Methodology, Software, Writing - Original Draft, Validation, Formal analysis, and Investigation. \textbf{G. G. Piva:} Conceptualization, Methodology, Software, Writing - Original Draft, Validation, Formal analysis, and Investigation. \textbf{A. S. da Mata:} Conceptualization, Writing - Review \& Editing, Supervision, Project administration, and Funding acquisition. \textbf{F. L. Ribeiro:} Conceptualization, Writing - Review \& Editing, Supervision, Project administration, and Funding acquisition.

\section*{Funding}
A.S. da Mata acknowledges the financial support from CNPq/Fapemig APQ-06591-24. 
F.\ L.\ Ribeiro thanks CNPq (grant numbers 403139/2021-0 and 424686/2021-0) and Fapemig (grant number APQ-00829-21
and APQ-06541-24) for financial support.

\end{document}